\documentclass{osa-article}

\journal{osajournal}

\articletype{Research Article}

\usepackage{verbatim}
\usepackage{physics}

\begin{document}

\title{Near perfect two-photon interference out a
down-converter on a silicon photonic chip}

\author{Romain Dalidet\authormark{1 $\dagger$}, Florent Mazeas\authormark{2 $\dagger$}, Edgars Nitiss\authormark{2}, Ozan Yakar\authormark{2}, Anton Stroganov\authormark{3}, Sébastien Tanzilli\authormark{1}, Laurent Labonté\authormark{1}, and Camille-Sophie Brès\authormark{2,*}}

\address{\authormark{1}Université Côte d’Azur, CNRS, Institut de Physique de Nice, 06108 Nice Cedex 2, France\\
\authormark{2}Ecole Polytechnique F\'ed\'erale de Lausanne, Photonic Systems Laboratory (PHOSL), CH-1015, Switzerland\\
\authormark{3}Ligentec SA, Lausanne Switzerland

\authormark{$^{\dagger}$}The two authors contribute equally to this work}
\email{\authormark{*}camille.bres@epfl.ch} 



\begin{abstract*}
Integrated entangled photon-pair sources are key elements for enabling large-scale quantum photonic solutions, and addresses the challenges of both scaling-up and stability. Here we report the first demonstration of an energy-time entangled photon-pair source based on spontaneous parametric down-conversion in silicon-based platform through an optically induced second-order ($\chi^{(2)}$) nonlinearity, ensuring type-0 quasi-phase-matching of fundamental harmonic and its second-harmonic inside the waveguide. The developed source shows a coincidence-to-accidental ratio of 1635 at 8 of $\mu$W pump power. Remarkably, we report two-photon interference with near-perfect visibility of 99.36$\pm1.94\%$, showing high-quality photonic entanglement without excess background noise. This opens a new horizon for quantum technologies requiring the integration of a large variety of building functionalities on single chips.

\end{abstract*}

 \vspace{7mm} 
Quantum science benefits from a worldwide research effort for developing disruptive technique finding repercussion in communication and processing of information. The realization of flexible, scalable and operational devices in quantum technologies and systems requires new approaches for hardware components. Integrated quantum photonics addresses the challenges of both scaling-up and stability, enabling the realization of key physical resources, such as on-chip photon pair generators, filters, modulators and detectors, and especially their interconnects, leading the quest for advanced large-scale monolithic devices~\cite{noauthor_rise_2020, tanzilli_genesis_2012}. The range of desired building functions, each performing a special task, has motivated the emergence of hybrid platforms combining the best of different photonic technologies, including $\chi^{(2)}$ and $\chi^{(3)}$ nonlinearities based platforms~\cite{alibart_quantum_2016, zhu_integrated_2021} in single functional chips~\cite{meany_hybrid_2014, elshaari_hybrid_2020,silverstone_silicon_2016, oser_high-quality_2020}. The price to pay lies in the complex hybrid integration of each elementary building blocks revealing new technological issues, such as excessive optical losses due to transitions \cite{kaur_hybrid_2021}. An elegant solution, still keeping additional benefits of merging $\chi^{(2)}$- and $\chi^{(3)}$-nonlinearities based platforms, without compromising the overall functionality of the single device, lies in inducing an effective $\chi^{(2)}$ nonlinearity in a silicon (Si)-based chip.

Among multiple physical approaches, the Si-based technology remains an outstanding option due to ubiquitous capabilities developed by the microelectronics industry. While the linear circuitry in Si photonics is well established, achieving flexible, efficient and integrated quantum sources that would allow entangled photon-pair generation has remained challenging\cite{harris_integrated_2014, oser_high-quality_2020}. The latter is essential for further advances of photonic quantum devices and applications \cite{Caspani2017}. The creation of entangled photons in Si-based devices is naturally obtained through spontaneous four-wave mixing (SFWM), which generates photonic noise, degrading the overall system’s performance. Recent publications show that the ability to detect photon correlations (entanglement) is limited by spurious effects such as two-photon and free-carrier absorption (TPA and FCA)~\cite{rosenfeld_mid-infrared_2020}. Furthermore, restrictive phase-matching conditions limit the spectral coverage since the entangled photon pairs are created only over a few tens of nanometers apart from the pump wavelength. In contrast, $\chi^{(2)}$ materials represent a viable alternative by leveraging spontaneous parametric down-conversion (SPDC) and thus relaxing the stringent condition for on-chip filtering as the generated photon pairs are spectrally far separated from the pump. Previous works have already shown the possibility of enabling an efficient $\chi^{(2)}$ response in a $\chi^{(3)}$ Si substrate providing new grounds for exploiting SPDC on this platform~\cite{castellan_origin_2019, billat_large_2017, nitiss_optically_2021,Porcel:17, timurdogan_electric_2017}. All these works have reported only either technological or non-linear characterizations. Also, temporal correlation measurements for revealing the degree of indistinguishability between the paired photons have been performed on aluminium nitride~\cite{guo_parametric_2017}.
In addition to demonstrating SPDC, qualifying quantum correlations in a rigorous way, \textit{i.e.}, by means of a Bell-type entanglement witness, is mandatory for a large variety of quantum applications such as secret key distribution\cite{ekert_quantum_1991}, superdense coding\cite{mattle_dense_1996}, teleportation~\cite{de_riedmatten_long_2004}, sensing\cite{brus_distributed_2004}, and computing\cite{qiang_large-scale_2018}. Such an entanglement characterization still remains unrealized for photon pairs generated via SPDC on a Si-based platform.

In this work, we demonstrate and qualify a photon-pair source obtained via SPDC process on an Si-based platform. The source relies on an effective $\chi^{2}$ induced by all-optical poling in a stoichiometric silicon nitride (Si$_3$N$_4$) waveguide \cite{nitiss_formation_2020}, resulting in the inscription of a self-organized grating enabling quasi-phase-matching (QPM) of $1560~{\rm nm}$ and $780~{\rm nm}$ light in a second-harmonic (SH) generation process. The spectrum out of the chip is fully telecom compliant, and spreads over the C- and L-bands. Low noise performance is demonstrated by measuring high coincidence-to-accidental ratios (up to 1500). We proceed to the qualification of energy-time entanglement carried by the photon pairs using a standard Franson-type interferometer. Remarkably, we show two-photon interference fringes with a near perfect raw visibility exceeding 99.36$\pm1.94\%$. To our knowledge, this work stands as the first demonstration of entangled photons out of an Si-based platform through SPDC, and opens new horizon for these platforms in the framework of quantum technologies.

\begin{figure}[ht]
\centering\includegraphics[width = 7 cm]{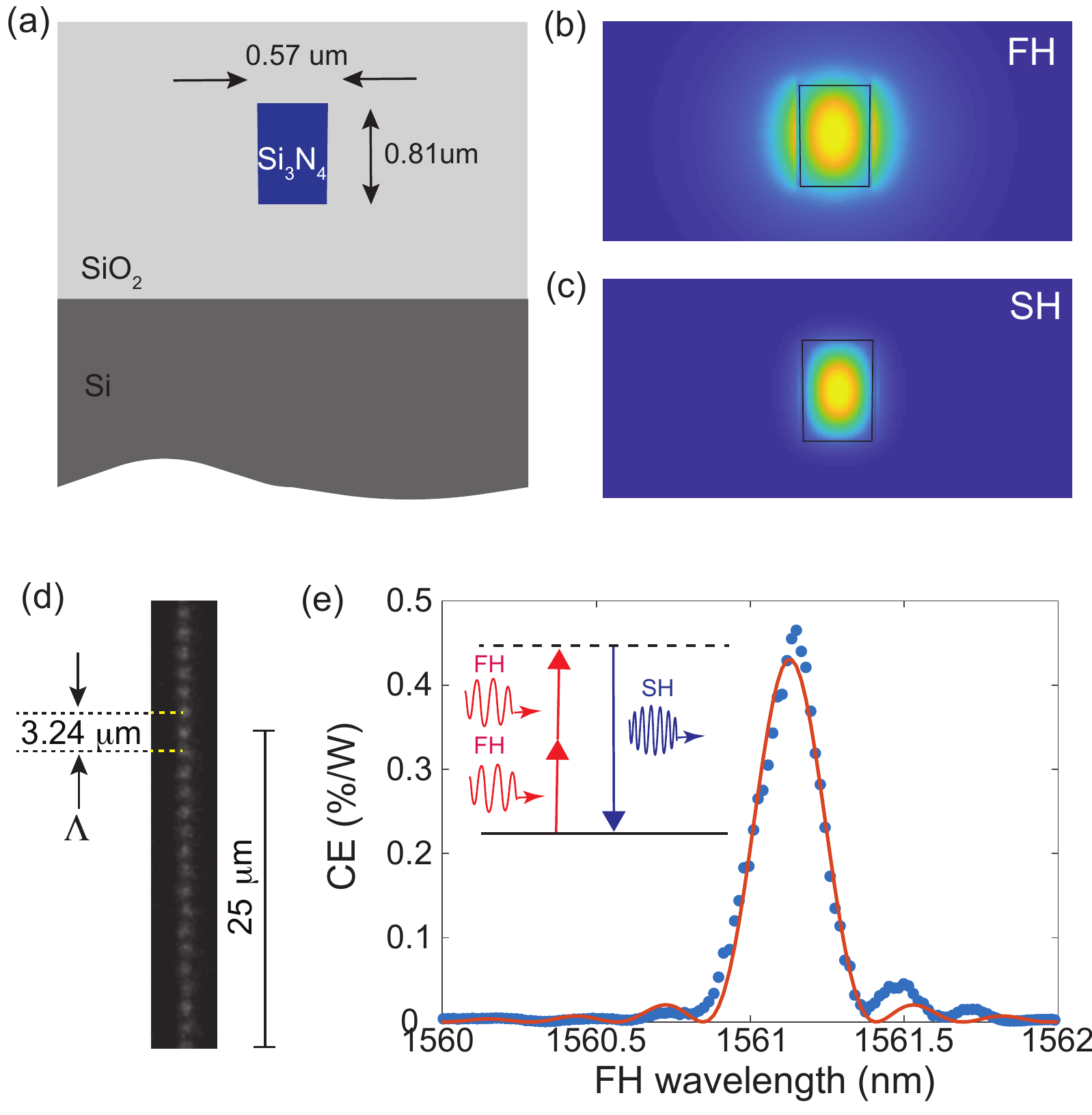}
\caption{\textbf{a} Cross-section of the Si$_3$N$_4$ waveguide (see also a top-view in Fig. \ref{Figure2}(a). \textbf{b} Simulated TE$_{\rm 00}$ mode at the FH near 1550 nm. \textbf{c} Simulated TE$_{\rm 00}$ mode at SH near 775 nm. \textbf{d} Effective refractive indices. \textbf{e} TPM image of the waveguide after all optical poling at 1561 nm showing the inscribed grating. \textbf{f} SH generation CE spectrum as a function of FH wavelength. Dots are data points and line is a fit according to Eq. \eqref{QPM}.
}
\label{Figure1} 
\end{figure}

We obtain the generation of entangled photons at $1560~{\rm nm}$ from a pump at $780~{\rm nm}$ in a $81~{\rm mm}$ long Si$_3$N$_4$ waveguide with a cross-section of $0.57\times0.81~{\rm \mu m^2}$ folded in meanders on a chip (Fig. \ref{Figure1}a). The coupling of light to and off the chip is implemented by inverse tapers.
We measure coupling losses for $1560~{\rm nm}$ and $780~{\rm nm}$ light to be $2.7~{\rm dB}$ and $4.6~{\rm dB}$ per facet, while the propagation loss is $0.33~{\rm dB/cm}$ and $0.73~{\rm dB/cm}$, respectively. The fundamental TE$_{00}$ modes for $1560~{\rm nm}$ and $780~{\rm nm}$ light are shown in Fig. \ref{Figure1}b and c, respectively. The $\chi^{(2)}$ in the waveguide is achieved using an all-optical poling technique as described in ref~\cite{billat2017large}. All-optical poling relies on high peak power ns pulses coupled to the waveguide which, in the presence of weak SH, enables the spontaneous build-up of a charge-separated $\chi^{(2)}$ grating due to photogalvanic effect\cite{dianov1995photoinduced,anderson1991model}. The latter ensures automatic type-0 QPM of the fundamental harmonic (FH) $1560~{\rm nm}$ and SH $780~{\rm nm}$ waves during the grating writing process, and high versatility such that the process can occur for a wide range of poling periods and waveguides indifferent of dimensions \cite{NitissACS}. For this study we all-optically poled the waveguide at $1561~{\rm nm}$ for operation in TE mode. Based on the simulated effective refractive indices, the criterion for type-0 QPM between the TE$_{00}$ modes at both the fundamental and SH waves requires a grating with small period $\Lambda = 3.14~{\rm \mu m}$, owing to the large wavevectors mismatch. All-optical poling automatically results in the inscription of such grating as confirmed by two-photon microscope (TPM) image shown in Fig. \ref{Figure1}e.
The shape of self-organized $\chi^{(2)}$ grating is directly linked to the mixing of the FH and SH modes along the waveguide. The latter results in an optimal overlap integral of the profiles of optical modes and the nonlinearity throughout the waveguide length \cite{NitissACS}.

Once the grating is inscribed, it is characterized by measuring the SH generation spectrum. The obtained on-chip conversion efficiency (CE) for an SH generation process as a function of the wavelength is shown in Fig. \ref{Figure1}f. Here CE is defined as $CE=P_{SH}/P^2_{P}$ where $P_{SH}$ and $P_{P}$ are the on-chip SH and pump powers, respectively. The CE spectra is fitted using the standard QPM function\cite{fejer1992quasi,wang2018ultrahigh}:
\begin{equation}
    {\rm CE}= \frac{(\omega L_{\rm g} \chi^{(2)}_{\rm eff})^2}{2\epsilon_0 c^3 n^2_{\rm P} n_{\rm SH}}
    \frac{S_{\rm SH}}{S^2_{\rm P}}
    \left(\frac{{\rm sin}\left(\frac{\Delta k L_{\rm g}}{2}\right)}{\left(\frac{\Delta k L_{\rm g}}{2}\right)}\right)^2,
\label{QPM}
\end{equation}
where $\omega$ is the FH optical frequency, $L_{\rm g}$ the grating length, $c$ the speed of light in vacuum, $\epsilon_0$ the vacuum permittivity, and $\Delta k$ the net wavevector mismatch after QPM compensation, i.e. ${\Delta k (\omega)}={k_{\rm SH}}-{2k_{\rm P}}-2{\pi}/{\Lambda}$. $k_{\rm P (SH)}$, $n_{\rm P(SH)}$, and $S_{\rm P (SH)}$ are the propagation constants, effective refractive indices, and the effective mode areas at FH and SH wavelength, respectively. Here $S_{\rm P}=0.74~{\rm \mu m^2}$ and $S_{\rm SH}=0.32~{\rm \mu m^2}$. By fitting the CE spectrum we extract the grating length, grating period and effective second-order nonlinearity to be $L_{\rm g}=69~{\rm mm}$, $\Lambda=3.14~{\rm \mu m}$ and $\chi^{(2)}_{\rm eff}=0.05~{\rm pm/V}$, respectively. The value of the grating period extracted from TPM image is $\Lambda_{\rm TPM}=3.24\pm0.31~{\rm \mu m}$ which is in excellent agreement with the one obtained from fitting of the CE data and with the theoretically expected value, confirming the fundamental mode interaction between the FH and the generated SH.

\begin{figure}[ht]
\centering\includegraphics[width = \linewidth]{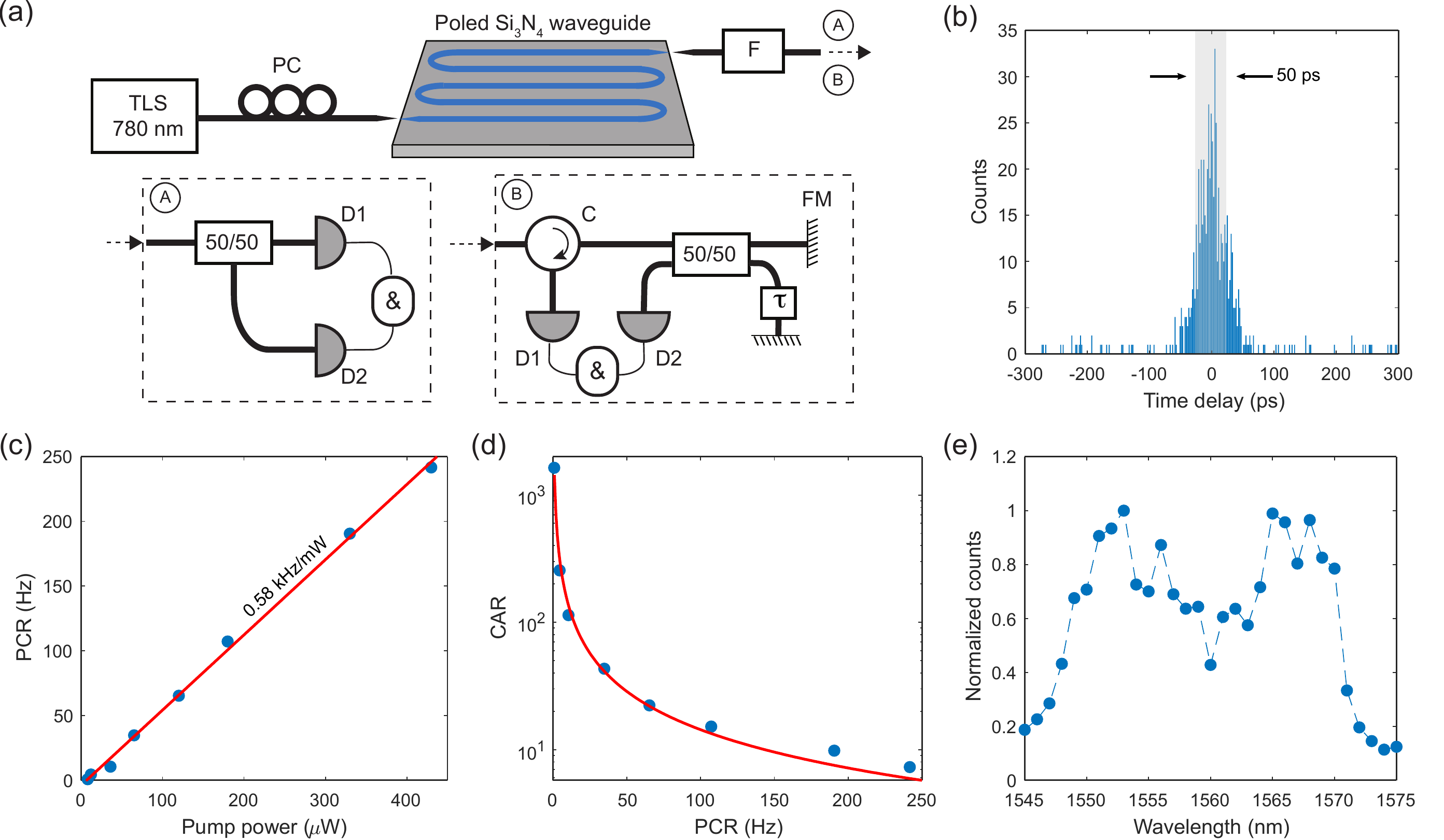}
    \caption{\textbf{a} Experimental setup. The output of the optically poled Si$_3$N$_4$ waveguide is connected to either (A) a coincidence measurement setup or (B) to the entanglement qualification setup in a folded-Franson configuration. The interferometer is an unbalanced Michelson made of long and short arms. PC: polarization controller; F: filter and isolators; 50/50: fiber coupler; $\tau$ time delay; D: superconducting nanowire single-photon detector; $\&$: coincidence counting electronics; FM: fiber Faraday mirror; C: circulator. \textbf{b} Coincidence count histogram in case pump power is 36 $\rm \mu W$.
    \textbf{c} Measured PCR produced by SPDC as a function of the coupled pump power.
    \textbf{d} Measured CAR as a function of PCR. Dots are data points, red lines are fits. The inset shows the count histogram at pump power of 36$\rm \mu W$. The coincidence counts $C_C$ are collected within a 80 ps time window
    \textbf{e} Normalized SPDC coincidence spectrum of the all-optically poled Si$_3$N$_4$ waveguide under test. Dashed line is a guide for the eye.
    }
\label{Figure2}
\end{figure}

The waveguide with the inscribed grating is then used for SPDC and characterization in the quantum regime, as shown in the experimental setup of Fig. \ref{Figure2}a. A $780~{\rm nm}$ continuous-wave pump laser is coupled to the all-optically poled Si$_3$N$_4$ waveguide. The polarisation of the pump laser is adjusted using a polarisation controller and aligned with the TE mode of the waveguide. At the output, the pump laser is rejected using a spectral filter. The produced twin photons are then sent to either a 50/50 fiber coupler, followed by two superconducting nanowire single-photon detectors (SNSPDs) (part A), or to a folded Franson consisting of a single unbalanced Michelson fiber interferometer (part B) in order to experimentally acquire two-photon interference fringes.

We exploit energy-time entangled photons as information carriers due to their inherent invulnerability against polarization mode dispersion and drifts~\cite{tittel_quantum_2000}, making this approach more robust for long distance distribution~\cite{aktas_entanglement_2016}.
A signature of energy-time entangled photon pairs is obtained by acquiring a coincidence histogram between the single-photon detectors (D1 and D2) as shown in Fig. \ref{Figure2}b. To characterize the correlated photon pairs, we measure the coincidence counts ($C_C$) and accidental-coincidence counts ($A_C$) as a function of the pump power. The $C_C$ is collected within a 50 ps time window covering the peak of the count histogram. The $A_C$ is estimated by collecting the total number of coincidence counts outside the central peak over a time window of 240 ps, much greater than the one of the central peak, before normalisation. The external pair coincidence rate (PCR) is calculated for several values of the coupled pump power and the results are shown in Fig. \ref{Figure2}c together with a fitted line whose slope is 0.58 kHz/mW. We now consider the internal brightness of SPDC-based sources, which is generally described by the slope efficiency given as the photon-pair flux per unit spectral width and per unit power, taking into account the overall losses of the setup. The overall coincidence spectrum, as will be detailed below, spreads over a spectral bandwidth of 30~nm, and shows an average of 400 coincidences per second with 14 dB of losses. We thus estimate an internal brightness of $\sim 5\cdot10^{-3}$ pairs/s/mW/MHz.

The measured coincidence-to-accidental ratio (CAR = $C_C/A_C$) with respect to PCR is shown in Fig. \ref{Figure2}d. For low pump powers ($<$1$\mu $W), we could not register accidental coincidences outside the coincidence peak within the measurement time of around 30 s. For a coupled pump power of 8 $\mu$W, we measure a CAR equal to 1635, which is a clear signature of correlated events without the presence of extra photonic noise. At low pump powers, the CAR is mainly limited by the detector’s dark counts. Then, the CAR decreases with the pump power, following the trend CAR $\propto$ PCR$^{-1}$ as shown in Fig. \ref{Figure2}d. The CAR drops because the pair rate increases quadratically with the pump power, yielding an increased probability of detecting multiple pairs within a given time window. To our knowledge, CAR obtained at low pump powers is the best value reported to date in the literature for entangled photon pairs emitted from Si-based platforms\cite{Clemmen:09,Silverstone:14,Wakabayashi:15,Steidle:15,Zhang:16,Wang:18,Samara:19,Choi:20}. This confirms our strategy of considering SPDC process in CMOS photonics platform.

Fig. \ref{Figure2}e shows the spectrum of the biphoton state out of the Si$_3$N$_4$ waveguide measured via non-local filtering by placing a variable filter in one arm of the setup, exploiting energy correlation of the photon pairs. The spectrum centered at 1560~nm spreads over part of both the C- and L-bands, thus covering 25 telecom channels. This flexibility, combined with the potential to center the spectrum at will as leveraged by all-optical poling, is particularly appealing for exploiting higher dimensions available in a large multiplexed telecom-channel spectrum~\cite{reimer_generation_2016}.


\begin{figure}[h!]
\centering\includegraphics[width =  \linewidth]{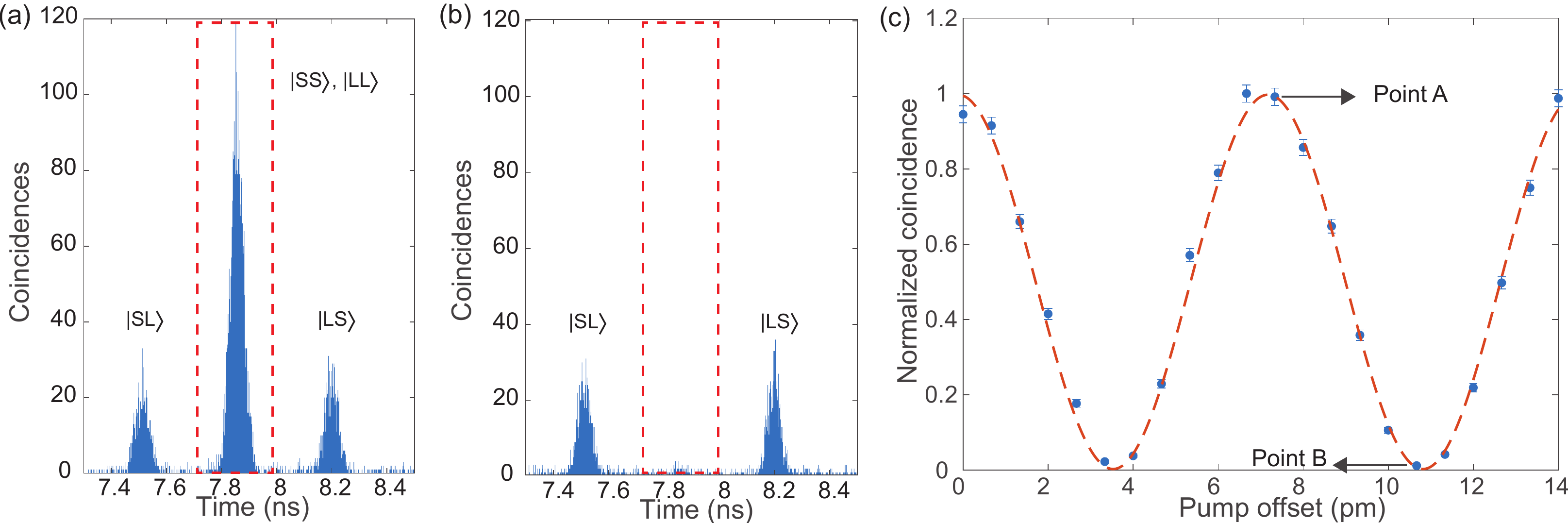}
\caption{\textbf{a} Coincidence histogram for the constructive interference setting at a pump offset of 7.35 pm (point A) from initial wavelength of 779.75 nm. \textbf{b} Coincidence histogram for the destructive interference setting at a pump offset of 10.65 pm (point B). Red-color dashed rectangle: analyzers time window. \textbf{c} Normalized interference fringes of energy-time entangled photons as a function of the pump offset. Error bars correspond to the Poissonian distribution inherent to the pump laser. From fit the obtained visibility is 99.36$\pm$1.94 \%.} 
\label{Figure3}
\end{figure}

The photon pairs are genuinely energy-time entangled as they are produced by SPDC~\cite{aktas_entanglement_2016}. To measure the quality of the produced entanglement, we use a setup as shown in Fig. \ref{Figure2}a part B. The photons pairs are directly sent to the folded-Franson interferometer and a typical three-peak coincidence histogram is recorded (Fig~\ref{Figure3} a and b). The contribution to these peaks depends on the path taken by the two photons. If the twin photons travel both via different arm, the time delays being $\tau_{short-long}$ ($|SL\rangle$) and $\tau_{long-short}$ ($|LS\rangle$) give rise to the two side peaks, arriving earlier and later, respectively. However, if the twin photons take both the same optical path, then, whether it is the long arm or the short arm, the time delays $\tau_{short-short}$ and $\tau_{long-long}$ ($|SS\rangle$ and $|LL\rangle$) will contribute to the central peak in an indistinguishable way.
The inability to distinguish the "which-path information" for these coincidences, is at the origin of the two-photon interference. In order to achieve high visibility two-photon interference fringes, our analysis system fulfils two requirements. First, the difference of time travel between the two arms is greater than the coherence time of the single photons to avoid single-photon interference $\tau_{sp}\approx 270fs < \tau \approx 30 ps) $. Then, this time travel difference is shorter than the coherence time of the CW pump laser $(\tau_{pump} > 10\mu s)$ to guarantee a coherent superposition between the indistinguishable states recorded in the central peak. 

We quantify the amount of entanglement by varying the pump wavelength offset and measuring the interferogram. As expected, the height of the side peaks remains the same during the scan, as they correspond to distinguishable events (see Fig. \ref{Figure3}a and b). The height of the central peak contains two (indistinguishable) contributions ($|SS\rangle$) and ($|LL\rangle$), instead of the side peak which gathers only one (distinguishable) contribution ($|SL\rangle$) or ($|LS\rangle$), leading a double amplitude, in average, for the central peak with respect to the side ones. Due to constructive interference (Fig. \ref{Figure3}a), the central peak amplitude increases by a factor 4 with respect to the side peak, whereas, destructive interference (Fig. \ref{Figure3}b) cancels any coincidence counts within the time analyzer window (Fig. \ref{Figure3}b). The normalized height of the central peak as a function of pump offset is plotted in Fig. \ref{Figure3}c. As expected, the trend is well fitted by a sinusoidal, revealing a near-ideal visibility of 99.36$\pm$1.94 \% without background subtraction, thus demonstrating the high quality of both the entanglement generator and associated measurement setup. Note that all results are well above the threshold of $71\%$ required to violate adapted Bell-Clauser-Horne-Shimony-Holt inequalities. This result stands as the first demonstration of correlated photon pairs through SPDC in CMOS compliant photonics platform, but also among the highest raw quantum interference visibility for time-energy entangled photons from a nanoscale integrated circuit.

In conclusion, here we demonstrate, for the first time to our knowledge, the realization of an entangled photons-pair source on Si-photonics platform based on SPDC, standing among the lowest-noise photon pair generators on a nanophotonic chip. The device relies on an effective $\chi^{(2)}$ induced by all-optical poling in a Si$_3$N$_4$ waveguide that enables near-perfect visibility entangled photon generation. Our device, whose fabrication process fulfils the requirements in terms of flexibility, compactness and scalability, could open a new horizon in the ongoing quest of monolithic photonic platforms compliant with the stringent demands of most quantum applications.

\begin{backmatter}

\bmsection{Funding}
This work was supported in part by ERC grant PISSARRO (ERC-2017-CoG 771647).
\bmsection{Disclosures}
The authors declare no conflicts of interest.

\end{backmatter}



\end{document}